\documentclass[prl,twocolumn,showpacs,preprintnumbers,amsmath,amssymb]{revtex4}

\usepackage{natbib}
\usepackage{bibentry}
\usepackage{graphicx}% Include figure files
\usepackage{dcolumn}% Align table columns on decimal point
\usepackage{bm}% bold math

\newcommand{\ket}[1]{|#1\rangle}
\begin{document}

\preprint{APS/123-QED}

\title{Entanglement Trapping in Structured Environments}

\author{Bruno Bellomo$^1$}
\author{Rosario Lo Franco$^1$}
\email{lofranco@fisica.unipa.it}
 \homepage{http://www.fisica.unipa.it/~lofranco}
\author{Sabrina Maniscalco$^2$}
\author{Giuseppe Compagno$^1$}
\affiliation{$^1$CNISM and Dipartimento di Scienze Fisiche ed Astronomiche,
Universit\`{a} di Palermo, via Archirafi 36, 90123 Palermo, Italy\\
$^2$Department of Physics, University of Turku, Turun yliopisto, FIN-20014 Turku, Finland}

\date{\today}% It is always \today, today,

\begin{abstract}
The entanglement dynamics of two independent qubits each embedded in a structured environment under conditions of inhibition of spontaneous emission is analyzed, showing entanglement trapping. We demonstrate that entanglement trapping can be used efficiently to prevent entanglement sudden death. For the case of realistic photonic band-gap materials, we show that high values of entanglement trapping can be achieved. This result is of both fundamental and applicative interest since it provides a physical situation where the entanglement can be preserved and manipulated, e.g. by Stark-shifting the qubit transition frequency outside and inside the gap.
\end{abstract}


\maketitle
Entanglement preservation is an important challenge in quantum information and computation technologies \cite{nielsenchuang}. Realistic quantum systems are affected by decoherence and entanglement losses because of the unavoidable interaction with their environments \cite{petru}. For example in Markovian (memoryless) environments, in spite of an exponential decay of the single qubit coherence, the entanglement between two qubits may completely disappear at a finite time \cite{diosi,yu2004PRL}. This phenomenon, known as "entanglement sudden death" and proven to occur in a quantum optics experiment \cite{almeida2007Science}, in turn limits the time when entanglement could be exploited for practical purposes. It is therefore of interest to examine the possibility to preserve entanglement. In the case of environments with memory (non-Markovian), such as imperfect cavities supporting a mode resonant with the atomic transition frequency, revivals of two-qubit entanglement have been found \cite{bellomo2007PRL,bellomo2008PRA,maniscalco2008PRL}. These revivals, although effectively extending the possible usage time of entanglement, decrease with time and eventually disappear after a certain critical time. Moreover, when the qubits interact with a common environment, it has been shown that entanglement can be preserved by means of the quantum Zeno effect \cite{maniscalco2008PRL}.

In this paper we continue the investigation on physical systems and physical effects that may lead to effective long time entanglement protection. Since entanglement evolution and population decay have been previously shown to be related \cite{bellomo2008PRA}, one is led to investigate situations where population trapping occurs. This can happen in structured environments where the density of states presents a dip which can inhibit spontaneous emission in the region of the dip \cite{garraway1997PRA}. Among realistic physical situations, this effect is known to occur in photonic band-gap (PBG) materials \cite{yablonovitch1987PRL}. Entanglement can be generated when a pair of atoms near-resonantly coupled to the edge of a PBG present direct dipole-dipole interaction \cite{niko2002JModOpt}. On the other hand, a way to produce entangled independent atoms in PBG materials is to consider a three-dimensional photonic crystal single-mode cavity with a sufficiently high-quality $Q$ factor where Rydberg atoms can freely travel through the connected void regions \cite{guney2007JOptSocAm}. The atoms exchange photons with the cavity, represented by a single defect mode of the crystal resonant with the atomic transition frequency, which acts as an atomic entanglement catalyst. In this situation, the atoms can be considered independent because their transition frequency is well inside the PBG and their relative distance is large enough to make the dipole-dipole interaction negligible.

The main aim of this work is to analyze if and to what extent entanglement initially present may be preserved for a general class of states of noninteracting pairs of atoms. Finally, we also analyze how such a system may be realistically manipulated to perform quantum computation and quantum information processing.

Our system consists of an entangled couple of identical independent two-level atoms (qubits), with states $\ket{0},\ket{1}$ and transition frequency $\omega_0$, each embedded in a PBG material at zero temperature. The atoms are assumed to be sufficiently separated in order to neglect the dipole-dipole interaction. Under this condition it has been shown that the two-qubit dynamics can be obtained by the knowledge of that of the single qubit \cite{bellomo2007PRL}. Each atom is coupled to the surrounding radiation field in a three-dimensional periodic dielectric \cite{john1990PRL}. In this case, the Hamiltonian of the single atom-environment system can be put in the form \cite{john1994PRA}
\begin{equation}\label{HamiltonianIP}
\hat{H}=\hbar \sum_k \left[
 ( \omega_k-\omega_0) \hat{b}_k^\dag \hat{b}_k+
 ig_k\left(\hat{\sigma}_-\hat{b}_k^\dag-\hat{\sigma}_+\hat{b}_k\right)\right],
\end{equation}
where $\omega_0$ is the transition frequency and $\sigma_ \pm$ are the raising and lowering
operators of the qubit, $b_k^\dag $, $b_k $ are the creation and annihilation operators and $g_k$ the coupling
constant of the mode $k$ with frequency $\omega_k$
\begin{equation}\label{coupling constants}
    g_k=\frac{\omega_0d}{\hbar}\left(\frac{\hbar}{2\epsilon_0\omega_kV}\right)^{1/2} \mathbf{e}_k\cdot \mathbf{u}_d.
\end{equation}
Here $d$ and $\mathbf{u}_d$ are the absolute value and unit vector of the atomic dipole moment, $V$ is the sample volume,  $\mathbf{e}_k\equiv \mathbf{e}_{k,j} $ are the two transverse (polarization) unit vectors, and $\epsilon_0$ the Coulomb constant. The Hamiltonian of Eq.~(\ref{HamiltonianIP}) is physically equivalent to the standard Hamiltonian of a two-level atom interacting with the radiation field and it is obtainable from the last one by unitary transformations \cite{barnbook}.

We consider here a model of electromagnetic waves in a three-dimensional periodic dielectric where the photon dispersion relation $\omega_k$ is chosen to be isotropic. By symmetrizing $\omega_k$, one produces PBGs at the spheres $|\mathbf{k}|=m\pi/L$ $(m=1,2,\ldots)$, where $L$ is the lattice constant. In such ideal photonic crystals, a PBG is the frequency range over which the local density of electromagnetic states and the decay rate of the atomic population of the excited state vanish. Near the band-gap edges the density of states becomes singular \cite{john1990PRL}, the atom-field interaction becomes strong and one can expect modifications to the spontaneous emission decay. In terms of the detuning $\delta=\omega_0-\omega_c$ between the atomic transition frequency $\omega_0$ and the band edge frequency $\omega_c$, the atomic population of the excited state is given by $P(t)=P(0)|q(t)|^2$ where the quantity $q(t)$ may be written as \cite{john1994PRA}
\begin{multline}\label{equforq}
q(t)=2a_1x_1 e^{\beta x_1^2t+i\delta t}+a_1(x_2+\sqrt{x_2^2})e^{\beta x_2^2t+i\delta t} \\-\sum^3_{j=1}a_j\sqrt{x_j^2}\left[1-\Phi \left(\sqrt{\beta x_j^2t}\right)\right]e^{\beta x_j^2t+i\delta t},
\end{multline}
where $\beta^{3/2}=\omega_0^{7/2}d^2/6\pi\epsilon_0\hbar c^3$, $\Phi(x)$ is the error function and $a_j=x_j/[(x_j-x_i)(x_j-x_k)]$ $(j\neq i\neq k;j,i,k=1,2,3)$ with
\begin{eqnarray}\label{xfunctions}
x_1&=&\left(A_+ + A_-\right)e^{i(\pi/4)},\nonumber\\
x_2&=&\left(A_+e^{-i(\pi/6)}+A_-e^{i(\pi/6)}\right)e^{-i(\pi/4)},\nonumber\\
x_3&=&\left(A_+e^{i(\pi/6)}+A_-e^{-i(\pi/6)}\right)e^{i(3\pi/4)},\nonumber
\end{eqnarray}
and $A_\pm=[1/2\pm\sqrt{1+4\delta^3/27\beta^3}]^{1/3}$. In the case when the environment is at zero temperature and the atom is initially in a general superposition state of its two levels, the single-qubit density matrix evolution can be expressed as \cite{petru}
\begin{equation}\label{roS}
\hat{\rho}^S(t)=\left(%
\begin{array}{cc}
\rho^S_{11}(0)|q(t)|^2  & \rho^S_{10}(0)q(t)\\\\
\rho^S_{01}(0)q^*(t)  & \rho^S_{00}(0)+ \rho^S_{11}(0)(1-|q(t)|^2) \\
\end{array}\right).
\end{equation}
The expression for $q(t)$ of Eq.~(\ref{equforq}) thus allows to obtain the single qubit dynamics according to Eq.~(\ref{roS}) and then to determine the two-qubit dynamics by exploiting a procedure that relies on the knowledge of the single-qubit reduced dynamics \cite{bellomo2007PRL}. In particular, in the standard basis $\mathcal{B}=\{\ket{1}\equiv\ket{11},
\ket{2}\equiv\ket{10}, \ket{3}\equiv\ket{01}, \ket{4}\equiv\ket{00}
\}$, the diagonal elements of the two-qubit reduced density matrix $\hat{\rho}(t)$ at time $t$ are
\begin{eqnarray}\label{rototdiag}
\rho_{11}(t)&=&\rho_{11}(0)|q(t)|^4,\nonumber\\
\rho_{22}(t)&=&\rho_{11}(0)|q(t)|^2(1-|q(t)|^2)+\rho_{22}(0)|q(t)|^2,\nonumber\\
\rho_{33}(t)&=&\rho_{11}(0)|q(t)|^2(1-|q(t)|^2)+\rho_{33}(0)|q(t)|^2,\nonumber\\
\rho_{44}(t)&=&1-[\rho_{11}(t)+\rho_{22}(t)+\rho_{33}(t)],
\end{eqnarray}
and the non-diagonal elements
\begin{eqnarray}\label{rototnodiag}
\rho_{12}(t)&=&\rho_{12}(0)q(t)|q(t)|^2,\ \rho_{13}(t)=\rho_{13}(0)q(t)|q(t)|^2,\nonumber\\
\rho_{14}(t)&=&\rho_{14}(0)q(t)^2,\ \rho_{23}(t)=\rho_{23}(0)|q(t)|^2,\nonumber\\
\rho_{24}(t)&=&\rho_{13}(0)q(t)(1-|q(t)|^2)+\rho_{24}(0)q(t),\nonumber\\
\rho_{34}(t)&=&\rho_{12}(0)q(t)(1-|q(t)|^2)+\rho_{34}(0)q(t),
\end{eqnarray}
with $\rho_{ij}(t)=\rho^*_{ji}(t)$, $\hat{\rho}(t)$ being a hermitian matrix. Note that Eqs.~(\ref{rototdiag}) and (\ref{rototnodiag}) permit to obtain the two-qubit density matrix evolution for any initial state.

To analyze the two-qubit entanglement dynamics we use the concurrence $C$ \cite{wootters1998PRL}, which attains its maximum value 1 for maximally entangled states and vanishes for separable states. We consider as initial states the Bell-like states
\begin{eqnarray}
\ket{\Phi}=\alpha\ket{01}+e^{i\gamma}\beta\ket{10},\quad\ket{\Psi}=\alpha\ket{00}+e^{i\gamma}\beta\ket{11},\label{bellstates}
\end{eqnarray}
where $\alpha,\beta$ are real and $\alpha^2+\beta^2=1$. For these initial conditions the concurrence at time $t$ is given respectively by
\begin{eqnarray}\label{concBell-likestates}
C_\Phi(t)&=&2\mathrm{max}\{0,\alpha\sqrt{1-\alpha^2}|q(t)|^2\},\nonumber\\
C_\Psi(t)&=&2\mathrm{max}\{0,\sqrt{1-\alpha^2}|q(t)|^2\nonumber\\
&&\times[\alpha-\sqrt{1-\alpha^2}(1-|q(t)|^2)]\}.
\end{eqnarray}
For initial Bell (maximally entangled) states, $\ket{\Phi}=\left(\ket{01}\pm\ket{10}\right)/\sqrt{2}$ and $\ket{\Psi}=\left(\ket{00}\pm\ket{11}\right)/\sqrt{2}$, the concurrences at the time $t$ take the simple form
\begin{equation}\label{concurrenceBell}
C_\Phi(t)=|q(t)|^2,\quad C_\Psi(t)=|q(t)|^4,
\end{equation}
independently on the relative phase $\gamma$. The previous equation is relevant because it shows the direct link between the behavior of excited state population of single atom $P(t)$ and that of concurrence.

\begin{figure}
\begin{center}
\includegraphics[width=7.00 cm, height=4.8 cm]{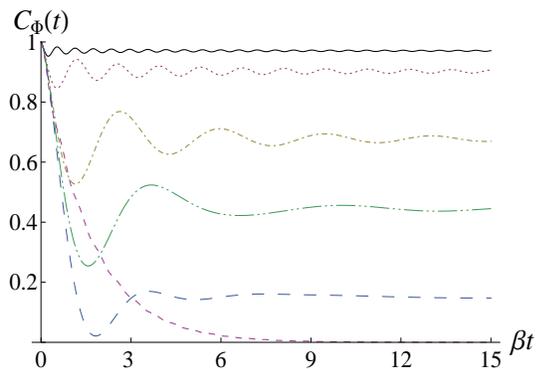}
\caption{\label{concurrencephi}\footnotesize{Concurrence $C_\Phi(t)$ as a function of $\beta t$
starting from the initial Bell-like state $\ket{\Phi}=\left(\ket{01}\pm\ket{10}\right)/\sqrt{2}$ for various values of detuning from photonic band edge:  $\delta = -10\beta$ (solid curve), $\delta = -4\beta$ (dotted curve), $\delta = -\beta$ (long-short-dashed curve), $\delta = 0$ (long-short-short-dashed curve), $\delta = \beta$ (long-dashed curve), $\delta = 10\beta$ (short-dashed curve).}}
\end{center}
\end{figure}
\begin{figure}
\begin{center}
\includegraphics[width=7.00 cm, height=4.8 cm]{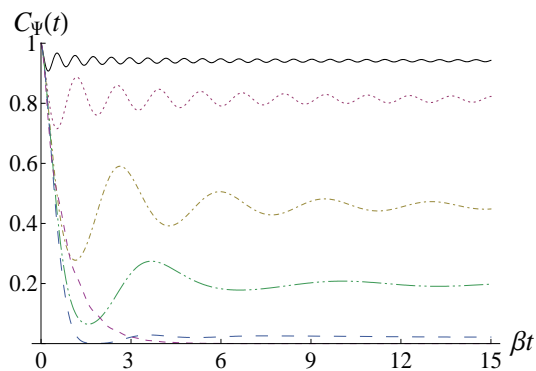}
\caption{\label{concurrencepsi}\footnotesize Concurrence $C_\Psi(t)$ as a function of $\beta t$
starting from the initial Bell-like state $\ket{\Psi}=\left(\ket{00}\pm\ket{11}\right)/\sqrt{2}$ for various values of detuning from photonic band edge:  $\delta = -10\beta$ (solid curve), $\delta = -4\beta$ (dotted curve), $\delta = -\beta$ (long-short-dashed curve), $\delta = 0$ (long-short-short-dashed curve), $\delta = \beta$ (long-dashed curve), $\delta = 10\beta$ (short-dashed curve).}
\end{center}
\end{figure}
\begin{figure}
\begin{center}
\includegraphics[width=8.00 cm, height=4.8 cm]{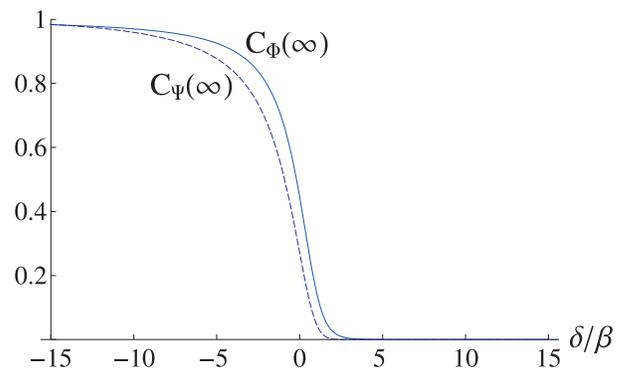}
\caption{\label{asymptoticvalues}\footnotesize Asymptotic values of concurrences $C_\Phi(t=\infty)$ (solid line) and $C_\Psi(t=\infty)$ (dashed line) as a function of $\delta/\beta $, starting respectively from the initial Bell state $\ket{\Phi}=\left(\ket{01}\pm\ket{10}\right)/\sqrt{2}$ and $\ket{\Psi}=\left(\ket{00}\pm\ket{11}\right)/\sqrt{2}$.}
\end{center}
\end{figure}

The concurrences $C_\Phi(t)$ and $C_\Psi(t)$ of Eq.~\ref{concurrenceBell}, corresponding to the initial Bell states $\ket{\Phi}$ and $\ket{\Psi}$ respectively, are shown in Fig.~\ref{concurrencephi} and Fig.~\ref{concurrencepsi} as a function of the dimensionless time $\beta t$ and for various values of detuning $\delta$, $\delta<0$ corresponding to the case when the two-level atom frequency $\omega_0$ is inside the band gap. The figures display that entanglement trapping is possible. The concurrence is nearer to its maximum value 1 for atomic frequencies at larger distances from the band edge and deeper inside the gap. From Fig.~\ref{concurrencephi} and Fig.~\ref{concurrencepsi} no qualitative difference appears between the evolutions of concurrences for the two initial states $\ket{\Phi}$ and $\ket{\Psi}$. A small quantitative difference can be appreciated in the asymptotic values of concurrences $C_\Phi(\infty)$ and $C_\Psi(\infty)$, with $C_\Psi(\infty)$ being slightly lower than $C_\Phi(\infty)$. The asymptotic values of concurrences for $t\rightarrow\infty$ can be obtained by the steady-state atomic population in the excited state $|q_s|^2=\lim_{t\rightarrow \infty} |q(t)|^2=4|a_1x_1|^2$ \cite{john1990PRL} on the basis of Eq.~(\ref{concurrenceBell}). In Fig.~\ref{asymptoticvalues} $C_\Phi(\infty)$ and $C_\Psi(\infty)$ are plotted as a function of the ratio $\delta/\beta $. We see that, when $\delta/\beta<0$, these asymptotic values differ from zero and decrease very rapidly moving near the edge. However, also for a small range of $\delta/\beta>0$, they are not zero. This can be linked to the presence of a photon-atom bound dressed state in the single qubit dynamics \cite{john1990PRL}. It is evident that the more the atomic transition frequency is far from the band edge inside the gap ($\delta/\beta<0$), the higher is the preserved entanglement. This figure confirms that the preserved entanglement starting from the Bell state $\ket{\Psi}$ is smaller than that relative to the initial state $\ket{\Phi}$. As said, entanglement trapping is achievable even when $\omega_0$ lies outside the band gap near the edge, although with small asymptotic values ($<0.4$). Exploiting Eq.~(\ref{concBell-likestates}) it is possible to study the behavior of concurrence in terms of the initial degree of entanglement (represented by $\alpha$). The result is that, as one might expect, smaller initial entanglement leads to lower asymptotic value of concurrence. Entanglement sudden death and its consequent limit on the entanglement usage can be thus prevented in such systems.

The results here obtained are valid for ideal PBG materials, thus one may ask to what extent they hold for real crystals. In fact, in real crystals with finite dimensions a pseudogap is typically obtained where the density of states is strongly smaller than that of free space but is not exactly zero. In such gaps, however, the lowering of the local density of states can be so large that the spontaneous decay of an excited emitter is effectively inhibited so that memory and coherent control effects are admitted and similar to those occurring in an ideal PBG  \cite{sprik1996EPL,john1998PRE,john1999PRA}. This would correspond, in view of the relation between population and entanglement expressed by Eqs.~(\ref{concBell-likestates}) and (\ref{concurrenceBell}), to inhibition of entanglement decay.

Another aspect that must be considered is the possibility of performing, in the conditions here considered, two-qubit operations for quantum computing. When the qubit transition frequencies are inside the gap, the effect of any resonant external signal is impeded. Thus, a way to bring the qubit frequencies outside the gap is required for the time necessary to externally manipulate the qubits. This could be realized, e.g., by means of Stark shifting the frequencies with a static electric field. The duration of this external control should be short enough to limit the decay of entanglement. For example, suppose that two Rb Rydberg atoms, having transition frequency $\omega_0\sim50$ GHz \cite{hagley1997PRL}, are embedded in a photonic crystal built in such a way to have PBG frequencies in the range of GHz \cite{ozbay1994APL}. For such atoms the dipole moment is $d\sim2\times10^{-26}$ Cm \cite{guney2007JOptSocAm}, that in turn gives $\beta\sim20$ KHz, while the typical Stark shifts are $\Delta\sim200$ KHz \cite{hagley1997PRL}. This implies a dimensionless shift with respect to $\beta$ of the order $\Delta/\beta\sim10$. On the basis of the results displayed in Fig.~\ref{asymptoticvalues}, this shift is therefore large enough to move the atomic transition frequency, for example, from a ratio $\delta/\beta=-5$ inside the gap, corresponding to an entanglement trapping value $\sim0.9$, to a ratio $\delta/\beta=5$ outside the gap, where the atoms can be manipulated by an external signal.

In many experiments the qubits are mimicked by ``artificial atoms'' consisting in quantum dot structures embedded in the solid fraction of the PBG material \cite{john1999PRA,lodhal2004Nature}. Quantum dots permit the coherent manipulation of a single localized quantum system with the technological advantages of solid-state systems. In fact, the coherent control of an exciton wave function in a quantum dot, namely the manipulation of the relative phases of the eigenstates in a quantum superposition, is experimentally achievable \cite{bonadeo1998Science}.  Also in the quantum dots scenario one can tune the energy levels by means of the Stark effect with typical shifts $\sim$1--10 GHz \cite{alen2003APL,warburton2002PRB}, so that the required conditions for qubit coherent manipulations could be accomplished. These features make quantum dots incorporated in a PBG material good candidates for the implementation of various schemes for quantum computation and coherent information processing.

In the context of Rydberg atoms traveling through the connected void regions of a photonic crystal, opportune atom-cavity interactions allow to perform quantum logic gates, as dual-rail Hadamard and NOT gates \cite{guney2007JOptSocAm}, and also other quantum information procedures based on cavity QED seem to be achievable. After the gate operations, when the atoms are outside the cavity in the periodic region of the crystal (PBG), the entanglement between the two qubits can be preserved for a very long time due to the entanglement trapping phenomenon we have demonstrated in this paper.

In conclusion, in this paper we have presented a new phenomenon, namely entanglement trapping in structured reservoirs, and we have shown that it can be used to prevent entanglement sudden death. We have investigated the connection between entanglement trapping and the well known phenomenon of population trapping occurring in photonic band gap structures. We have shown that the entanglement, as measured by concurrence, initially present between independent qubits embedded in structured reservoirs, as PBG materials, can be preserved depending on the ``position'' of the qubit transition frequencies inside the band gap. The qubits may be coherently manipulated by an external control field to perform some quantum computation or information protocol. A possible way to achieve this is by a Stark shift that moves the qubit transition frequencies outside the gap for a time necessary to the desired two-qubit operation. We have pointed out that Rydberg atoms and quantum dots in photonic crystals may be suitable to this purpose. These considerations clearly highlight the potential of structured environments as a promising background in the development of new quantum technologies.

R.L.F. (G.C.) acknowledges partial support by MIUR project II04C0E3F3 (II04C1AF4E) \textit{Collaborazioni Interuniversitarie ed Internazionali tipologia C}. S.M. acknowledges financial support from the Academy of Finland (projects 108699, 115682), the Magnus Ehrnrooth Foundation and the V\"{a}is\"{a}l\"{a} Foundation.

\end{document}